\documentclass[twocolumn,amsmath,floatfix,amssymb,aps,prb,showpacs,letter,superscriptaddress]{revtex4-1}
\usepackage{graphicx,natbib}
\usepackage{color}

\begin{document}

\title{Temperature and Magnetic Field Dependence of Spin Ice Correlations in the Pyrochlore Magnet Tb$_{2}$Ti$_{2}$O$_{7}$}

\author{K. Fritsch}
\affiliation{Department of Physics and Astronomy, McMaster University, Hamilton, Ontario, L8S 4M1, Canada}
\author{E. Kermarrec}
\affiliation{Department of Physics and Astronomy, McMaster University, Hamilton, Ontario, L8S 4M1, Canada}
\author{K. A. Ross}
\affiliation{Institute for Quantum Matter and Department of Physics and Astronomy, Johns Hopkins University, Baltimore, Maryland 21218, USA}
\affiliation{NIST Center for Neutron Research, NIST, Gaithersburg, Maryland 20899-8102, USA}
\author{Y. Qiu}
\affiliation{NIST Center for Neutron Research, NIST, Gaithersburg, Maryland 20899-8102, USA}
\affiliation{Department of Materials Science and Engineering, University of Maryland, College Park, Maryland 20742, USA}
\author{J. R. D. Copley}
\affiliation{NIST Center for Neutron Research, NIST, Gaithersburg, Maryland 20899-8102, USA}
\author{D.~Pomaranski}
\affiliation{Department of Physics and Astronomy and Guelph-Waterloo Physics Institute, University of Waterloo, Waterloo, Ontario, Canada N2L 3G1}
\affiliation{Institute for Quantum Computing, University of Waterloo, Waterloo, Ontario, Canada N2L 3G1}
\author{J. B. Kycia}
\affiliation{Department of Physics and Astronomy and Guelph-Waterloo Physics Institute, University of Waterloo, Waterloo, Ontario, Canada N2L 3G1}
\affiliation{Institute for Quantum Computing, University of Waterloo, Waterloo, Ontario, Canada N2L 3G1}
\author{H. A. Dabkowska}
\affiliation{Brockhouse Institute for Materials Research, Hamilton, Ontario, L8S 4M1, Canada}
\author{B. D. Gaulin}
\affiliation{Department of Physics and Astronomy, McMaster University, Hamilton, Ontario, L8S 4M1, Canada}
\affiliation{Brockhouse Institute for Materials Research, Hamilton, Ontario, L8S 4M1, Canada}
\affiliation{Canadian Institute for Advanced Research, 180 Dundas St.\ W., Toronto, Ontario, M5G 1Z8, Canada}

\date{\today}

\begin{abstract}

We present a parametric study of the diffuse magnetic scattering at $\left(\frac{1}{2},\frac{1}{2},\frac{1}{2}\right)$ positions in reciprocal space, ascribed to a frozen antiferromagnetic spin ice state in single crystalline Tb$_2$Ti$_2$O$_7$. Our high-resolution neutron scattering measurements show that the elastic (-0.02 meV $<E<0.02$ meV) $\left(\frac{1}{2},\frac{1}{2},\frac{1}{2}\right)$ scattering develops strongly below $\sim275$ mK, and correlates with the opening of a spin gap of $\sim0.06-0.08$ meV over most of the Brillouin zone. The concomitant low-lying magnetic spin excitations are weakly dispersive and appear to soften near the $\left(\frac{1}{2},\frac{1}{2},\frac{1}{2}\right)$ wave vector at 80 mK. The nature of the transition at 275 mK has many characteristics of spin glass behavior, consistent with ac-susceptibility measurements. The application of a magnetic field of 0.075 T applied along the [1-10] direction destroys the $\left(\frac{1}{2},\frac{1}{2},\frac{1}{2}\right)$ elastic scattering, revealing the fragility of this short-range ordered ground state. We construct a refined $H$-$T$ phase diagram for Tb$_2$Ti$_2$O$_7$ and [1-10] fields which incorporates this frozen spin ice regime and the antiferromagnetic long-range order previously known to be induced in relatively large fields. Specific heat measurements on the same crystal reveal a sharp anomaly at $T_{\rm c}\sim$ 450 mK and no indication of a transition near $\sim 275$ mK.  We conclude that the higher temperature specific heat peak is not related to the magnetic ordering but is likely a signal of other, nonmagnetic, correlations.

\end{abstract}
\maketitle

\section{Introduction}

The search for experimental realizations of quantum spin liquid (QSL) and quantum spin ice (QSI) states is of great current interest \cite{BalentsSL,OnodaGenericQSI2012} due to the potential for exotic physics within these systems. Geometrically frustrated magnetic materials are central to the discovery of such exotic states.\cite{frustratedspinsystems} Besides materials based on quasi two-dimensional (2D) lattices such as herbertsmithite~\cite{Han2012,herbertsmithite} or the organic spin liquid candidate $\kappa$-(BEDT-TTF)$_2$Cu$_2$(CN)$_3$,\cite{organictriangular,Pratt2011} the three-dimensional (3D) pyrochlore lattice consisting of networks of corner-sharing tetrahedra, has been the focus of significant and sustained research efforts.\cite{BalentsSL,JasonRMP,BramwellGingras,Castelnovo2012,OnodaGenericQSI2012} The family of rare earth-based titanate pyrochlores has played a prominent role in these studies as many members of this family can be grown as large and pristine single crystals.\cite{Savary,KatePRX,Fennell2009,Morris2009}  Within this family, Tb$_2$Ti$_2$O$_7$ stands out as perhaps the least well understood, despite having been studied in this context for well over a decade.\cite{Gardner1999} In this material, the Tb$^{3+}$ magnetic moments that decorate the pyrochlore lattice are expected to display Ising-like easy axis anisotropy with spins pointing along the local $\left\langle111\right\rangle$ axes, either into or out of the tetrahedra, as a consequence of them being non-Kramers ions and their crystal field (CF) doublet ground state.\cite{GingrasPRB2000,MirebeauPRB2007,Zhangarxiv} Based on a negative Curie-Weiss temperature ($\Theta_{\rm CW}\sim$ -14 K) and this Ising-like anisotropy, mean-field theory predicts the classical ground state of Tb$_2$Ti$_2$O$_7$ to not be frustrated, and Tb$_2$Ti$_2$O$_7$ is, on the basis of Monte Carlo simulations,\cite{GingrasdenHertog} expected to order into an antiferromagnetic (AF) ${\bf Q}=0$ N\'{e}el ordered state around $\sim$ 1 K.\cite{GingrasPRB2000} Yet, this material fails to exhibit conventional long-range order (LRO) down to at least 50 mK as shown in very early studies by Gardner \textit{et al.}\cite{Gardner1999,Gardner2003} 

This puzzle has motivated two vigorously debated\cite{MolavianQSI2007,MolavianQSI2009,BruceCF,Bonville2013a} theoretical scenarios to account for Tb$_2$Ti$_2$O$_7$'s disordered ground state: a quantum spin ice (QSI) scenario\cite{MolavianQSI2007,MolavianQSI2009,Molavian2009} and a nonmagnetic singlet ground state.\cite{BonvilleSinglet2011} The QSI proposal introduces quantum dynamics to the problem through virtual fluctuations of the Tb$^{3+}$ ions between their ground state and excited CF doublets.  These quantum fluctuations reposition Tb$_2$Ti$_2$O$_7$ into the nearby spin ice regime within a generalized phase diagram appropriate to Ising-like pyrochlore magnets.\cite{GingrasdenHertog} The nonmagnetic singlet scenario assumes a splitting of the accidental CF doublet ground state of the Tb$^{3+}$ ions ($J=6$, non-Kramers) into two non-magnetic singlets through a Jahn-Teller-like symmetry lowering of the CF environment, thus producing a non-magnetic ground state. The latter scenario implies a non-cubic structure at low but finite temperatures in Tb$_2$Ti$_2$O$_7$.  To date, evidence for such a static structural distortion is lacking at all but very large magnetic fields.\cite{JacobXray2007,Ruff2010}

Recent experiments on Tb$_2$Ti$_2$O$_7$ have shown signatures of the proposed QSI state, most notably pinch point scattering\cite{Fennell2012} indicative of Coulombic correlations in spin ice, and signs of a magnetization plateau for magnetic fields along $\left[111\right]$ theoretically proposed by Molavian \textit{et al.},\cite{Molavian2009} although there is a diversity in the interpretation of this latter evidence.\cite{Lhotel2012,Baker2012,Legl2012,Yin2013,Sazonov2013,unpublished} Our recent neutron scattering measurements \cite{Fritsch2013} on single-crystalline Tb$_2$Ti$_2$O$_7$ at 70 mK have revealed short-range spin ice correlations at $\left(\frac{1}{2},\frac{1}{2},\frac{1}{2}\right)$ wave vectors which can be ascribed to a short-range antiferromagnetically ordered spin ice state with spin canting at an angle of $\sim$12$^\circ$ from the local $\left\langle111\right\rangle$ axes. These static correlations observed in the elastic scattering channel were shown to be very sensitive to both applied magnetic field and temperature, although parametric studies of the field and temperature dependence of this scattering were not performed. The static $\left(\frac{1}{2},\frac{1}{2},\frac{1}{2}\right)$ correlations were separated from low-lying inelastic magnetic excitations by a spin gap of $\sim$0.06-0.08 meV. Other neutron studies also observed diffuse scattering intensity at $\left(\frac{1}{2},\frac{1}{2},\frac{1}{2}\right)$ positions, however, these studies did not relate the scattering to a short-range antiferromagnetically ordered spin ice state.\cite{Fennell2012,PetitTTO2012} A subsequent study of non-stoichiometric Tb$_{2+x}$Ti$_{2-x}$O$_{7+y}$ by Taniguchi \textit{et al.} \cite{Taniguchi2013} also observed the appearance of a $\left(\frac{1}{2},\frac{1}{2},\frac{1}{2}\right)$ Bragg-like peak associated with quasi-long-range order at low temperatures in a polycrystalline sample which was lightly ``stuffed",  Tb$_{2+x}$Ti$_{2-x}$O$_{7+y}$ ($x=0.005$). 

In this paper, we report a detailed study of the temperature and field dependence of the previously observed elastic $\left(\frac{1}{2},\frac{1}{2},\frac{1}{2}\right)$ and associated low-lying inelastic scattering in Tb$_{2}$Ti$_{2}$O$_{7}$. We show that the elastic scattering appears upon lowering the temperature below $\sim$ 275 mK and that its appearance correlates with the opening of a spin gap within the low-lying magnetic spectral weight over most of the Brillouin zone. This magnetic spectral weight and its associated spin excitations exhibit weak dispersion and appear to soften somewhat at the ``quasi''-ordering wave vector $\left(\frac{1}{2},\frac{1}{2},\frac{1}{2}\right)$ and at the lowest temperature measured, $T = 80$ mK. The application of a magnetic field along $\left[\right.$1-10$\left.\right]$ leads to a destruction of the short-range ordered ground state by $\mu_0 H=0.075$ T.  We construct a new $H$-$T$ phase diagram for Tb$_2$Ti$_2$O$_7$ in a $\left[\right.$1-10$\left.\right]$ magnetic field, which encompasses both this new frozen spin ice state at very low fields and temperature, and the previously known long-range ordered, antiferromagnetic state which exists at relatively high fields and over a much larger range of temperatures.

\section{Experimental Details}

The single crystal sample of Tb$_{2}$Ti$_{2}$O$_{7}$ used for the neutron scattering measurements we present here is the same sample used in earlier studies\cite{Fritsch2013,Rule2006} and was grown using the optical floating zone technique at McMaster University.\cite{HannaOFZ,Gardnercrystalgrowth} Time-of-flight neutron scattering measurements were performed using the disk-chopper spectrometer DCS \cite{CopleyDCS} at the NIST Center for Neutron Research. Two different incident energies were employed; for lower energy-resolution measurements we used $E_i$=3.27 meV, giving an energy resolution of 0.1 meV, while the higher energy-resolution measurements we report used $E_i$=1.28 meV with a resulting resolution of 0.02 meV. The sample was carefully aligned with the $\left[\right.$1-10$\left.\right]$ direction vertical to within 0.5$^\circ$, such that the (H,H,L) plane was coincident with the horizontal scattering plane. Measurements were performed in a temperature range of $\sim80$ mK $< T \lesssim 650$ mK and at magnetic fields $\leq 0.2$ T.

Specific heat measurements were performed with a $^3$He/$^4$He dilution refrigerator at the University of Waterloo. The crystal used for neutron scattering measurements was sectioned into a 33.8 mg mass with dimension 2.6 x 2.6 x 1.3 mm$^{3}$ and this smaller single crystal was used in the heat capacity study. The relaxation method was employed with a thermal weak link of manganin wire, with conductance 5.0 x $10^{-7}$ J/K/s at 0.80 K. The resulting time constant was greater than 600 seconds at the highest temperature measured. The average step size for the relaxation measurement was 3.5\% of the nominal temperature, with a minimum equilibration time window of five times the thermal relaxation constant.

\section{Magnetic Field dependence}

\begin{figure}
\includegraphics[width=8cm]{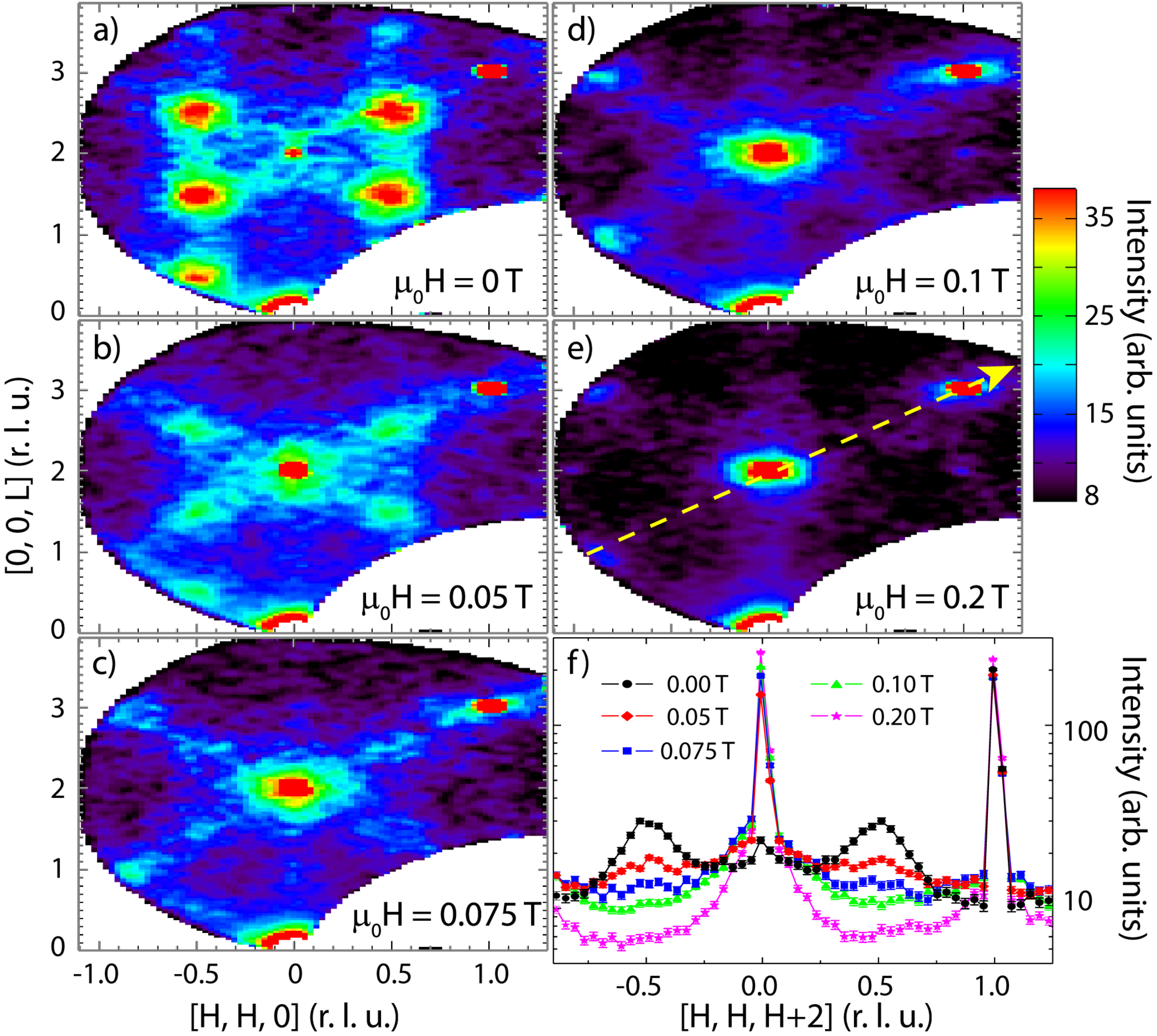}%
\caption[{Elastic neutron scattering data within the (H,H,L) plane of Tb$_{2}$Ti$_{2}$O$_{7}$ at $T = 80$ mK for different values of applied magnetic field ($H \parallel$ $\left[\right.$1-10$\left.\right]$). The energy is integrated from -0.1 meV $< E <$ 0.1 meV. Panels (a) through (e) show the evolution of the elastic scattering for an applied magnetic field of 0, 0.05, 0.075, 0.1 and 0.2 T, respectively. Panel (f) shows the {\bf Q}-dependence of the elastic scattering along the $\left[111\right]$ direction cutting through the $(0,0,2)$ Bragg position as indicated by the yellow arrow in panel (e).}]{Elastic neutron scattering data within the (H,H,L) plane of Tb$_{2}$Ti$_{2}$O$_{7}$ at $T = 80$ mK for different values of applied magnetic field ($H \parallel$ $\left[\right.$1-10$\left.\right]$). The energy is integrated from -0.1 meV $< E <$ 0.1 meV. Panels (a) through (e) show the evolution of the elastic scattering for an applied magnetic field of 0, 0.05, 0.075, 0.1 and 0.2 T, respectively. Panel (f) shows the magnetic field dependence of the elastic scattering along the $\left[111\right]$ direction cutting through the $(0,0,2)$ Bragg position as indicated by the yellow arrow in panel (e). The $\left(\frac{1}{2},\frac{1}{2},\frac{1}{2}\right)$ peaks vanish around $\mu_0H\sim$ 0.075 T. All data have an empty can background subtracted. The error bars are $\pm1\sigma$.}%
\label{fig:figure5b1}%
\end{figure}

Figure \ref{fig:figure5b1} shows a series of elastic scattering maps in the (H,H,L) plane for different magnetic fields applied along $\left[\right.$1-10$\left.\right]$ in panels (a) through (e). Here, elastic scattering integrates over -0.1 meV $< E <$ 0.1 meV in energy.  Figure \ref{fig:figure5b1}(f) shows a cut of this elastic scattering, along the $\left[111\right]$ high-symmetry direction covering the $(-0.5,-0.5,1.5)$, $(0.5,0.5,2.5)$ as well as the $(0,0,2)$ and $(1,1,3)$ Bragg positions, as indicated by the yellow dashed line in Fig. \ref{fig:figure5b1}(e). As the applied magnetic field is increased to 0.05 T, the previously observed strong diffuse scattering at the $\left(\frac{1}{2},\frac{1}{2},\frac{1}{2}\right)$ positions in zero field (panel (a)) gets suppressed substantially (panel (b)) and is much reduced by 0.075 T (panel (c)).

Simultaneous to the suppression of the $\left(\frac{1}{2},\frac{1}{2},\frac{1}{2}\right)$ diffuse scattering intensity in fields up to 0.075 T, we observe a strong build-up in magnetic elastic intensity at the structurally forbidden $(0,0,2)$ Bragg peak. As the field is increased further (panels (d) and (e)), most of the remaining diffuse scattering condenses into the Bragg position at $(0,0,2)$. The intensity scale is linear for Fig. \ref{fig:figure5b1}(a)-(e). In contrast, the cut along the $\left[111\right]$ direction through $(0,0,2)$ in Fig. \ref{fig:figure5b1}(f) is plotted on a logarithmic intensity scale. Note that the highest applied field here, 0.2 T, is still far below the critical $\left[\right.$1-10$\left.\right]$ field $\sim 2$ T, above which Tb$_2$Ti$_2$O$_7$ is found in a field-induced antiferromagnetic long-range ordered state, characterized by the appearance of resolution-limited spin waves and a strong $(1,1,2)$ magnetic Bragg peak.\cite{Rule2006}

A further feature in the elastic scattering maps are weak rods of scattering along both the $\left\langle00L\right\rangle$ and $\left\langle111\right\rangle$ directions that can be identified as evidence for significant anisotropic exchange in Tb$_2$Ti$_2$O$_7$.\cite{Fennell2012} Rods of magnetic scattering have been identified in other rare-earth titanate pyrochlores, such as Yb$_{2}$Ti$_{2}$O$_{7}$,~\cite{Ross2009} where anisotropic exchange is well established.\cite{KatePRX,Applegate2012,Hayre2013} Anisotropic exchange in Tb$_2$Ti$_2$O$_7$ has recently been investigated in papers by Bonville \textit{et al.}\cite{Bonville2013a} and by Curnoe,\cite{Curnoe2013} the latter of which related an effective $S=1/2$ spin Hamiltonian to diffuse magnetic neutron scattering.

\begin{figure}
\includegraphics[width=8cm]{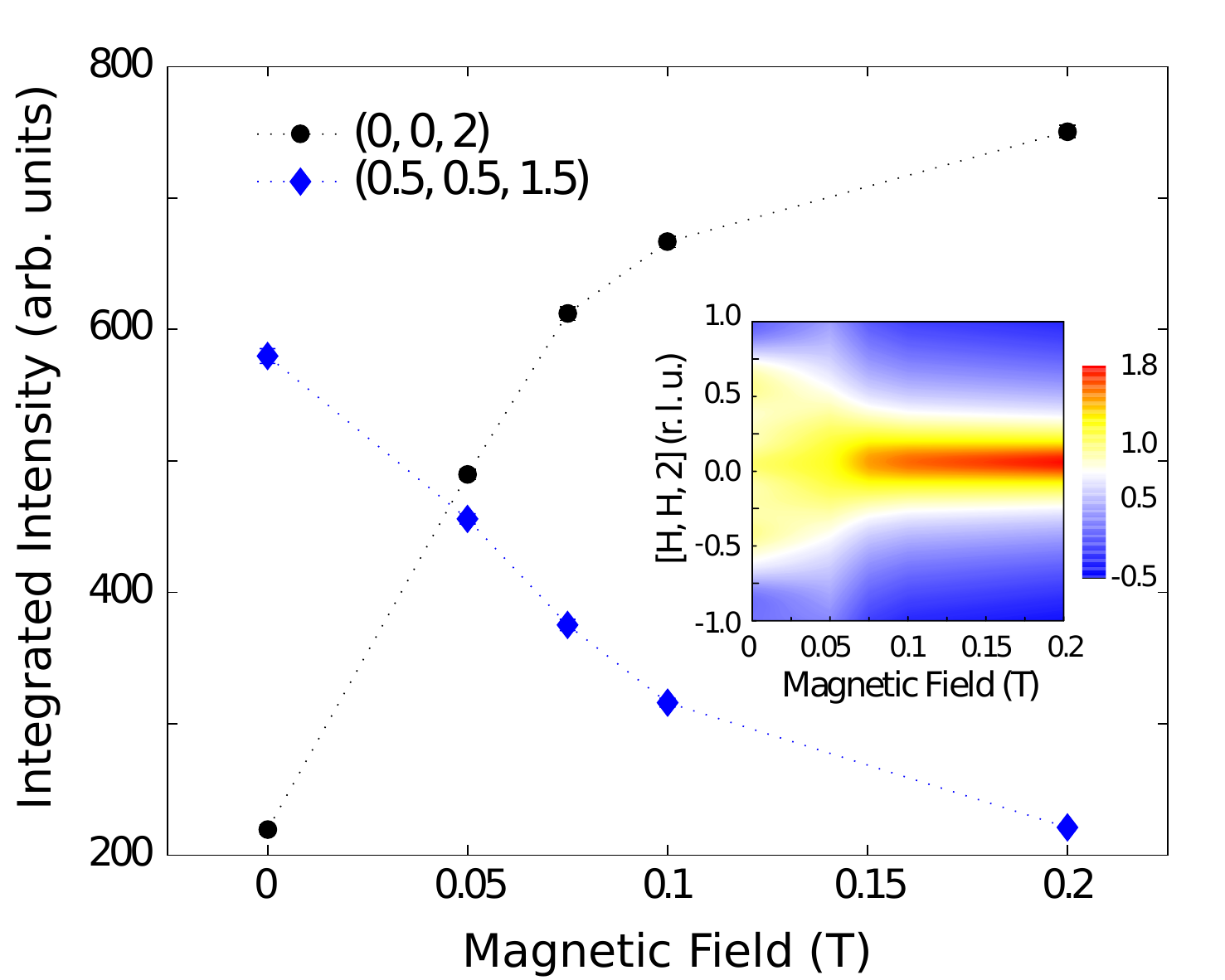}
\caption[{Integrated elastic scattering at the $\left(\frac{1}{2},\frac{1}{2},\frac{1}{2}\right)$ and $(0,0,2)$ positions as a function of magnetic field. The $\left(\frac{1}{2},\frac{1}{2},\frac{1}{2}\right)$ scattering gets clearly suppressed at $\sim0.075$ T. The inset shows a color contour map of the diffuse elastic scattering around the $(0,0,2)$ Bragg peak made up of line scans along [H,H,2].}]{Integrated elastic scattering at the $\left(\frac{1}{2},\frac{1}{2},\frac{1}{2}\right)$ and $(0,0,2)$ positions as a function of magnetic field. The $\left(\frac{1}{2},\frac{1}{2},\frac{1}{2}\right)$ scattering gets clearly suppressed at $\sim0.075$ T. The inset shows a color contour map of the diffuse elastic scattering around the $(0, 0, 2)$ Bragg peak made of line scans along [H,H,2] on a logarithmic intensity scale. The binning ranges for this cut were L$=[1.9, 2.1]$ r.l.u.. The intensity of the resolution-limited weak elastic scattering at (0,0,2) in zero field has been subtracted off. Diffuse scattering from the $\left(\frac{1}{2},\frac{1}{2},\frac{1}{2}\right)$ positions is seen to extend out to (0.5,0.5,2) and (-0.5,-0.5,2) below 0.075 T; above this field, the diffuse scattering increases in intensity and narrows considerably at the (0,0,2) position. An empty can background subtraction and correction for detector efficiency were performed for all data shown. Error bars are $\pm1\sigma$.}
\label{fig:figure5b2}
\end{figure}

One can integrate up the elastic scattering data in a relatively small region around  the $\left(\frac{1}{2},\frac{1}{2},\frac{1}{2}\right)$ or $(0,0,2)$ positions in reciprocal space, and look at its explicit $\left[\right.$1-10$\left.\right]$ magnetic field dependence. This is what is shown in Fig. \ref{fig:figure5b2}, with relatively small integration ranges of $\pm 0.2$ r.l.u. in [H,H,H], and $\pm0.15$ r.l.u. perpendicular to [H,H,H] for $\left(\frac{1}{2},\frac{1}{2},\frac{1}{2}\right)$, and $\pm0.1$ r.l.u. in [H,H,H] and $\pm0.15$ r.l.u. perpendicular to [H,H,H] for $(0,0,2)$. We identify the inflection point in the field dependence of $\left(\frac{1}{2},\frac{1}{2},\frac{1}{2}\right)$ near $\mu_0H\sim0.075$ T, as well as the leveling off of the $(0,0,2)$ elastic intensity for fields beyond 0.075 T with the $T\sim0$ phase boundary to the frozen, antiferromagnetic spin ice state in Tb$_2$Ti$_2$O$_7$. The inset to Fig. \ref{fig:figure5b2} shows a color map of the diffuse elastic scattering around the (0,0,2) position. This color map is made of a series of scans along [H,H,$2$], integrating over L$=[1.9,2.1]$ r.l.u.. The resolution limited weak scattering at (0,0,2) in zero magnetic field has been fit and removed from the scattered intensity shown in this inset. As evident from this map and consistent with the integrated elastic intensity shown in Fig. \ref{fig:figure5b2}, strong diffuse scattering at the $\left(\frac{1}{2},\frac{1}{2},\frac{1}{2}\right)$ positions is
present below $\mu_0H=0.075$ T, and this extends across these [H,H,2] scans, that is to (0.5,0.5,2) and (-0.5,-0.5,2). For fields above $\mu_0H=0.075$ T, this diffuse scattering collapses into the (0,0,2) position.

The field dependence of the elastic magnetic scattering, and in particular the rapid fall-off of the $\left(\frac{1}{2},\frac{1}{2},\frac{1}{2}\right)$ frozen spin ice elastic scattering near 0.075 T is in qualitative agreement with recent work by Yin \textit{et al.}.\cite{Yin2013} They have proposed a low-temperature low-field phase diagram for Tb$_2$Ti$_2$O$_7$ based on magnetization and ac-susceptibility data. These measurements reveal two magnetic phases, the first one of which appears below 140 mK and for fields $\mu_0H<0.067$ T, while another phase is found at very low temperatures $T< 40$ mK and higher fields $0.067<\mu_0H<0.6$ T. From the observation of a weak magnetization plateau in the presence of a $\left[111\right]$ magnetic field, theoretically predicted by Molavian \textit{et al.},\cite{Molavian2009} these authors conclude that a quantum kagome spin ice state exists at higher fields, with a QSI state at low and zero magnetic fields. Although these Yin \textit{et al.}\cite{Yin2013} measurements were performed in a $\left[111\right]$ applied magnetic field, we note that anisotropy in the magnetization of Tb$_2$Ti$_2$O$_7$ is only observed for field strengths exceeding $\gtrsim0.3$ T,\cite{unpublished} well above the range of applied field strengths which we report here in our neutron scattering measurements. For that reason, we consider the phase diagram presented in Ref. 32 to be germane to our findings, reinforcing our assertion that the $\left(\frac{1}{2},\frac{1}{2},\frac{1}{2}\right)$ elastic scattering is associated with a QSI state. 
\section{Temperature Dependence}

\begin{figure}
\includegraphics[width=8cm]{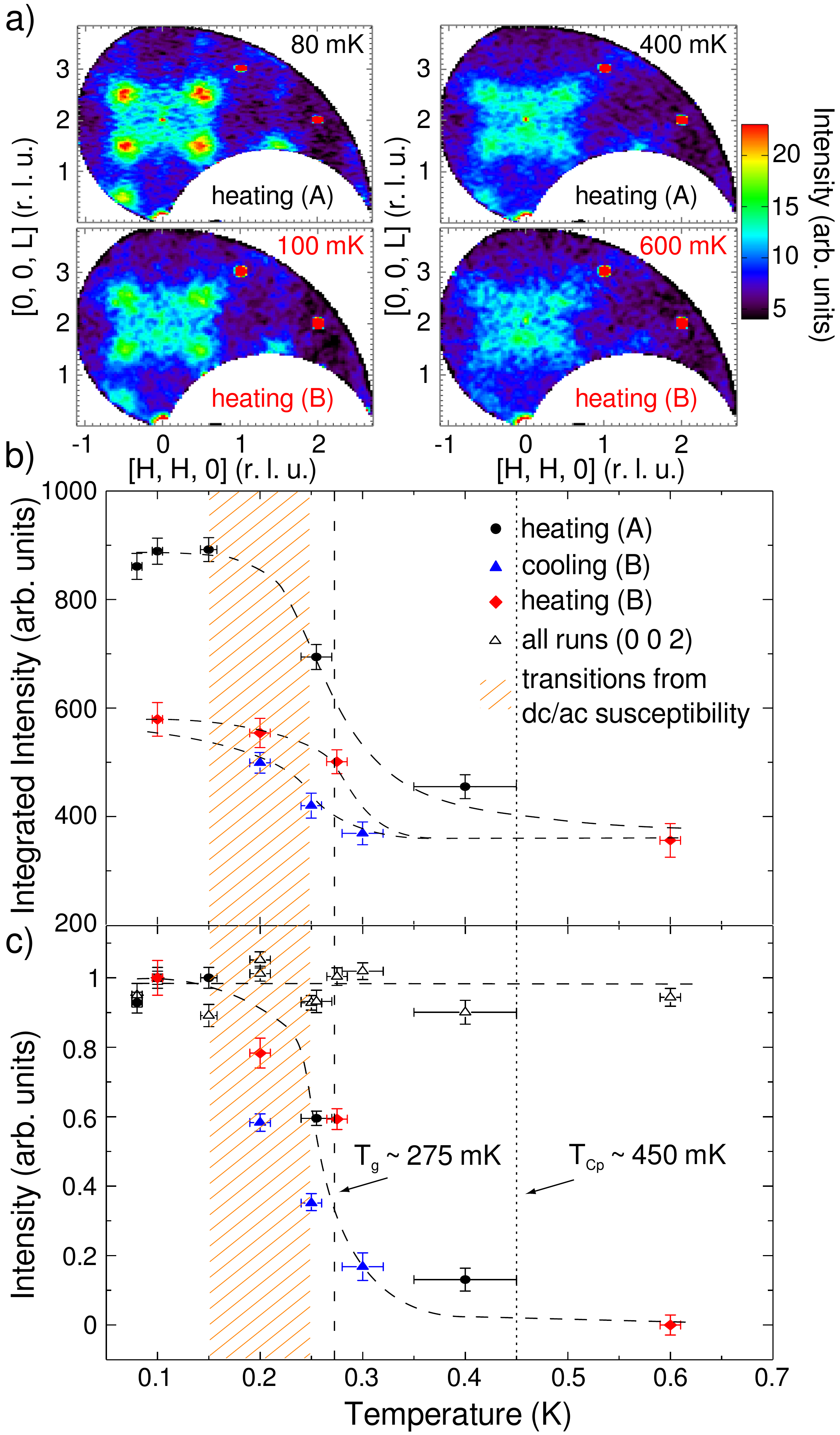}%
\caption[{Temperature dependence of the elastic $\left(\frac{1}{2},\frac{1}{2},\frac{1}{2}\right)$ scattering in zero field. The integrated elastic scattering intensity at the four $\left(\frac{1}{2},\frac{1}{2},\frac{1}{2}\right)$-like positions is presented as a function of temperature for different heating and cooling cycles in panel (b). Representative reciprocal space maps for the two heating cycles are shown in panel (a). Panel (c) shows the integrated intensity from (b) scaled to the difference between the low-temperature phase at 100 mK and the high-temperature phase at 600 mK (heating B) or 400 mK (heating A), placing the transition below which $\left(\frac{1}{2},\frac{1}{2},\frac{1}{2}\right)$ peaks appear at a temperature of $\sim275$ mK.}]{Temperature dependence of the elastic $\left(\frac{1}{2},\frac{1}{2},\frac{1}{2}\right)$ scattering in zero field. The integrated elastic scattering intensity ($-0.1$ meV $<E<0.1$ meV, H$=\pm\left[0.4, 0.6\right]$ r.l.u., L$=\left[1.4,1.6\right]$ and L$=\left[2.4,2.6\right]$ r.l.u.) at the four $\left(\frac{1}{2},\frac{1}{2},\frac{1}{2}\right)$-like positions is presented as a function of temperature for different heating and cooling cycles in panel (b). Please see text for details. Representative reciprocal space maps for the two heating cycles are shown in panel (a). Panel (c) shows the integrated intensity from (b) scaled to the difference between the low-temperature phase at 100 mK and the high-temperature phase at 600 mK (heating B) or 400 mK (heating A), placing the transition below which $\left(\frac{1}{2},\frac{1}{2},\frac{1}{2}\right)$ peaks appear at a temperature of $\sim275$ mK. In addition, the temperature dependence of the elastic $\left(0,0,2\right)$ scattering, normalized to the intensity at the lowest temperatures, is shown on the same plot, indicating no obvious temperature dependence. Various reports of this glassy transition from ac-susceptibility measurements\cite{Lhotel2012,Hamaguchi2004,Yaouanc2011,Gardner2003} are encompassed by the hatched area. The specific heat anomaly found for our crystal at $T_{\rm Cp}\sim450$ mK is indicated by a dotted vertical line. This anomaly appears at considerably higher temperature than the onset of the short-range magnetic ordering. Dashed lines serve as guide to the eyes.} 
\label{fig:figure5b3}
\end{figure}

The temperature dependence of the integrated elastic intensity of the $\left(\frac{1}{2},\frac{1}{2},\frac{1}{2}\right)$ peaks in zero field is presented in Fig. \ref{fig:figure5b3}. We determined the elastic integrated intensity (-0.1 meV $<$E$<0.1$ meV) at the four $\left(\frac{1}{2},\frac{1}{2},\frac{1}{2}\right)$ positions in our field of view in Fig. \ref{fig:figure5b3}(a) and average the integrated intensities at each temperature. The corresponding background signal for each $T$ was obtained by integrating over a region of low intensity in reciprocal space around $(1,1,2)$. This background term was subsequently subtracted from the averaged $\left(\frac{1}{2},\frac{1}{2},\frac{1}{2}\right)$ integrated elastic intensity. 

The results of these integrations are shown in Fig. \ref{fig:figure5b3}(b) for individual heating and cooling runs labeled in the order of the measurements. Two prominent features become apparent in Fig. \ref{fig:figure5b3}. First, the temperature dependence of the $\left(\frac{1}{2},\frac{1}{2},\frac{1}{2}\right)$ peaks behaves in an order parameter-like fashion with a distinctive drop in integrated intensity around 275 mK. Second, the ground state below 275 mK has some history dependence to it, with the elastic $\left(\frac{1}{2},\frac{1}{2},\frac{1}{2}\right)$ peak intensity variable from one thermal history to the next. History dependence is a common feature of spin glass states. \cite{Mydosh1993,BinderYoungReviewSG} We first discuss the second observation. Data points were collected in three temperature sweeps; heating A, cooling B and heating B. Representative elastic reciprocal space maps for the two runs obtained during heating are shown in Fig. \ref{fig:figure5b3}(a) for the low temperature short-range ordered ground state (80 mK or 100 mK) and for the higher temperature disordered state (400 mK or 600 mK, respectively). We observe that the absolute intensity of the ($\frac{1}{2}$,$\frac{1}{2}$,$\frac{1}{2}$) elastic peaks in the second heating run is a factor of $\sim$2 smaller, and that the overall diffuse elastic scattering background is somewhat larger compared to the first heating run. This indicates that a complex freezing process with several metastable states is likely at play at low temperatures. This is not so surprising, as the ($\frac{1}{2}$,$\frac{1}{2}$,$\frac{1}{2}$) elastic scattering does not correspond to long-range order, but rather is known to be characterized by a relatively short and isotropic correlation length of $\sim8$ \AA, again typical of spin glasses. Despite this, a common feature in all three heating and cooling runs is the appearance of the ($\frac{1}{2}$,$\frac{1}{2}$,$\frac{1}{2}$) peaks below 275 mK. To make that point, we scaled and normalized the integrated elastic intensities at ($\frac{1}{2}$,$\frac{1}{2}$,$\frac{1}{2}$) from Fig. \ref{fig:figure5b3}(b) to the difference in intensity between the low-temperature state at 100 mK or below and that of the high-temperature state (400 mK or 600 mK, respectively). The result is shown in Fig. \ref{fig:figure5b3}(c), which shows a glass-like transition to occur around $\sim$275 mK, independent of the detailed history dependence. We also note here that, in contrast to the elastic scattering at ($\frac{1}{2}$,$\frac{1}{2}$,$\frac{1}{2}$), the elastic scattering at (0,0,2) does not show any significant temperature dependence as shown in Figure \ref{fig:figure5b3}(c).

\begin{figure}
\includegraphics[width=8cm]{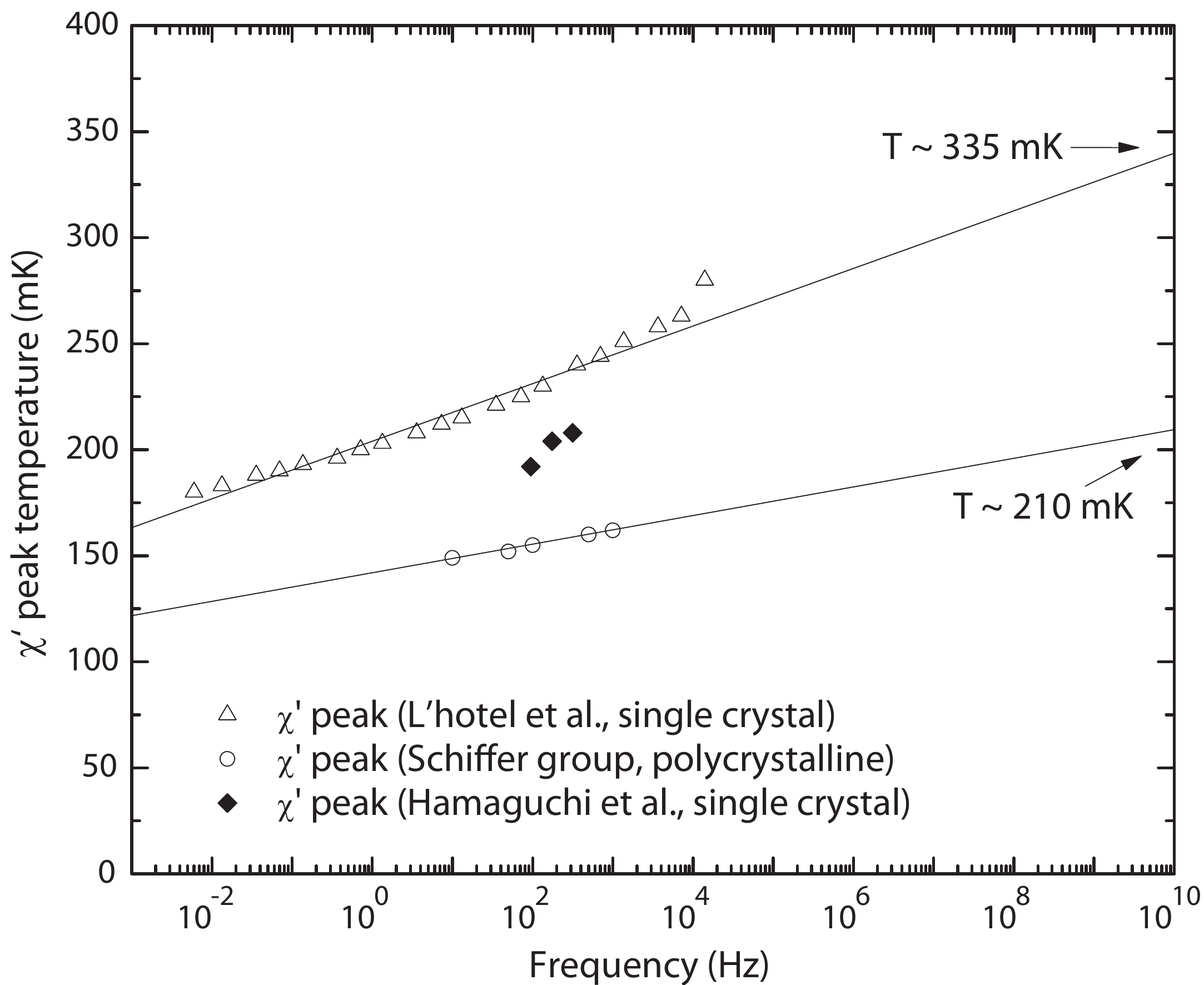}%
\caption[{Shift of the ac-susceptibility peaks per decade frequency for polycrystal and single crystalline samples from different groups.\cite{Lhotel2012,Hamaguchi2004,Schifferunpublished} For these three measurements, the slope $\frac{\Delta T}{T\Delta\left(log T\right)} <$0.1, typical of what is expected for spin glasses.\cite{Mydosh1993} The extrapolation of the ac-susceptibility peak to frequencies of the order 0.01 THz $\sim0.04$ meV energies, at which our measurements probe the system, shows that the expected transition occurs between 210-335 mK.}]{Shift of the ac-susceptibility peaks per decade frequency for polycrystal and single crystalline samples from different groups.\cite{Lhotel2012,Hamaguchi2004,Schifferunpublished} For these three measurements, the slope $\frac{\Delta T}{T\Delta\left(\log{T}\right)} <$0.1, consistent with expectations for insulating spin glasses.\cite{Mydosh1993} The extrapolation of the ac-susceptibility peak to frequencies of the order 0.01 THz $\sim0.04$ meV energies, at which our measurements probe the system, shows that the expected transition occurs between 210-335 mK. The transition temperature we determine falls right within that range and is thus consistent with these measurements.}
\label{fig:figure5b4}
\end{figure}

The spin glass transition as measured with the ($\frac{1}{2}$,$\frac{1}{2}$,$\frac{1}{2}$) elastic magnetic scattering is consistent with that observed by several groups using ac-susceptibility measurements. \cite{Lhotel2012,Hamaguchi2004,Gardner2003,Schifferunpublished} This data is summarized in Fig. \ref{fig:figure5b4}, and peaks found in $\chi '$, the real part of the ac-susceptibility, lie in the temperature range from $\sim$150-250 mK. This temperature regime is indicated with an orange hatched area in Fig. \ref{fig:figure5b3}(b) and (c), so that the results from elastic neutron scattering and ac-susceptibility can be compared. As neutron scattering measurements are performed at relatively high frequencies compared with ac-susceptibility, it is typical for neutron scattering to measure spin freezing associated with spin glass transitions at higher temperatures than are observed in ac-susceptibility. This is illustrated in Fig. \ref{fig:figure5b4}, in which we show the linearly extrapolated peak-shift per decade frequency, a phenomenological parameter used to describe the temperature dependence of the $\chi '$ peak in spin glasses as a function of frequency, to the frequency appropriate to our neutron measurements, $\sim$0.01~THz. Thus, a transition between 210-335 mK would be consistent with the range of previously reported AC susceptibility peaks
for different samples, which is in good agreement with our observations. This frequency dependence may indeed play a role here, although as seen in Fig. \ref{fig:figure5b3}(b) and (c), this does not seem to be a large effect as the agreement between the two techniques is quite good. We also note that Lhotel \textit{et al.}\cite{Lhotel2012} report a distinction between between the glassy behavior observed in Tb$_2$Ti$_2$O$_7$ and canonical spin glass behavior based on the frequency dependence of their ac-susceptibility data.  

\begin{figure}[h]
\includegraphics[width=8cm]{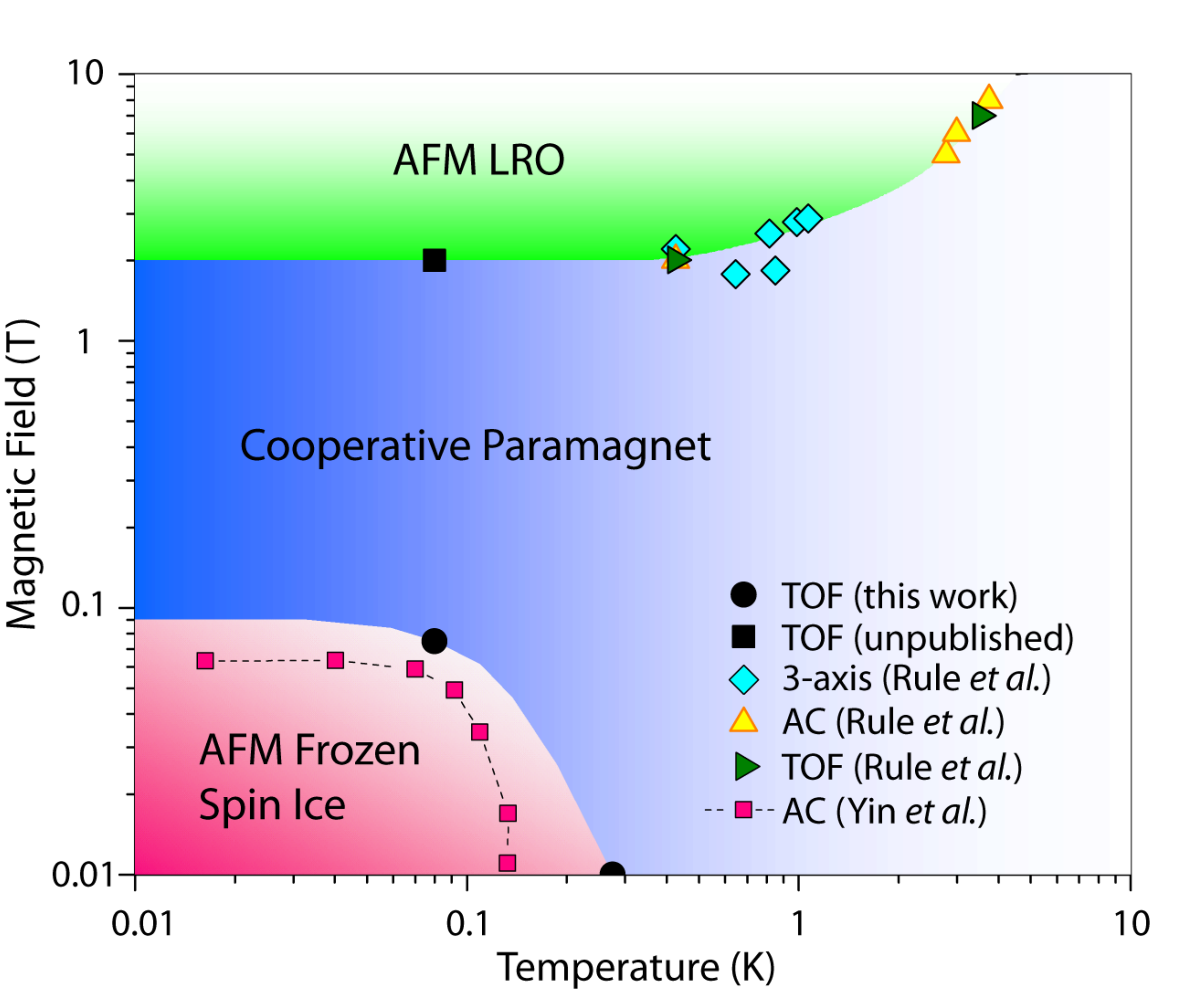}
\caption[{$H$-$T$ phase diagram for Tb$_2$Ti$_2$O$_7$ in [1-10] field extracted from neutron scattering measurements in the low field and low temperature region presented in this work and from high field and higher temperature measurements by Rule \textit{et al.}.\cite{Rule2006} The magnetic phase transitions measured by Yin \textit{et al.}~\cite{Yin2013} from ac-susceptibility with fields along [111] are also shown.}]{$H$-$T$ phase diagram for Tb$_2$Ti$_2$O$_7$ in [1-10] field extracted from neutron scattering measurements in the low field and low temperature region presented in this work and from high field and higher temperature measurements by Rule \textit{et al.}.\cite{Rule2006} The magnetic phase transitions measured by Yin \textit{et al.}~\cite{Yin2013} from ac-susceptibility with fields along [111] are also shown.}
\label{fig:figure5b5} 
\end{figure}

Based on the temperature and field dependence of the $\left(\frac{1}{2},\frac{1}{2},\frac{1}{2}\right)$ elastic scattering, we can construct an $H$-$T$ phase diagram for Tb$_2$Ti$_2$O$_7$ in [1-10] field, shown in Figure \ref{fig:figure5b5}, including the high-field and high-temperature regions determined by Rule \textit{et al.}'s work.\cite{Rule2006} Note that both temperature and field axes are on a logarithmic scale.

\begin{figure}
\includegraphics[width=8cm]{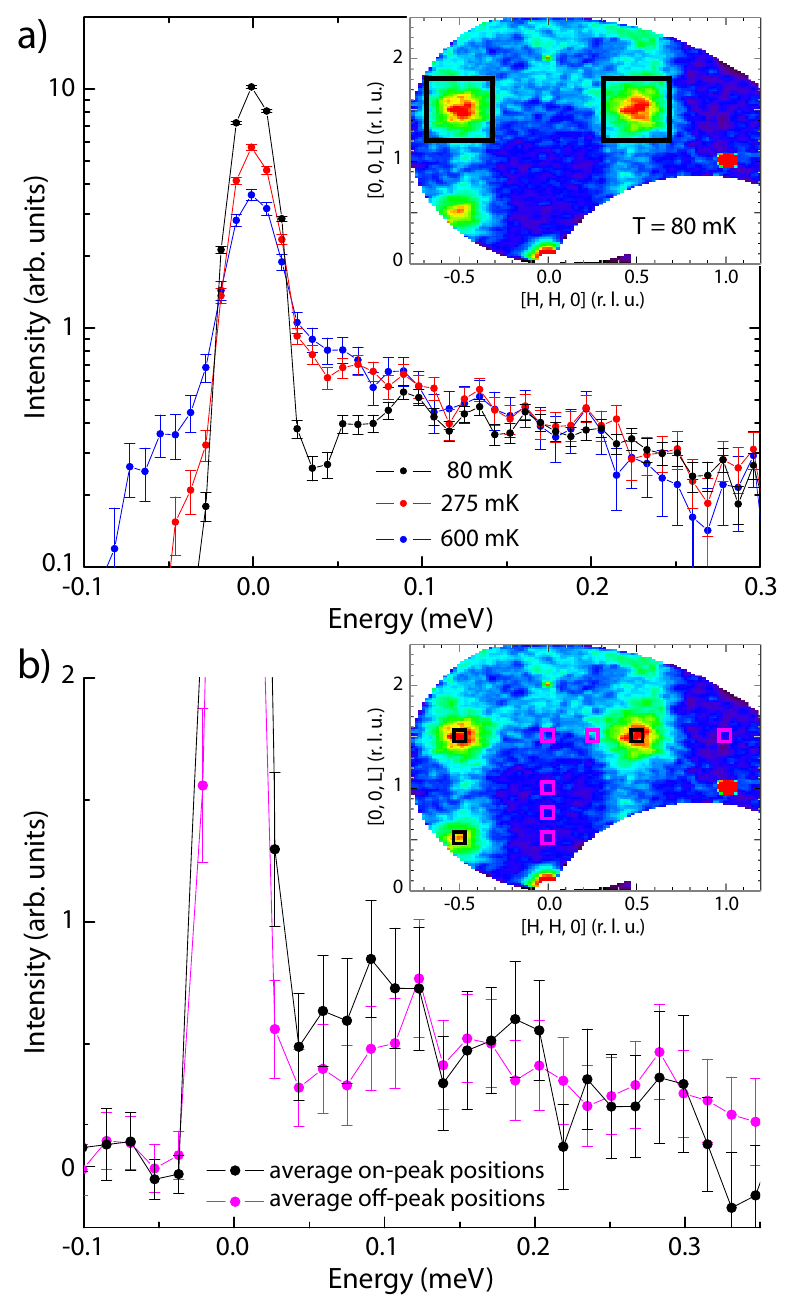}%
\caption[{High-resolution neutron scattering data of Tb$_2$Ti$_2$O$_7$ in zero field. Panel (a) shows a plot of intensity vs energy transfer of the ($\frac{1}{2}$,$\frac{1}{2}$,$\frac{1}{2}$) peak intensity averaged over two peaks as a function of temperature. Panel (b) shows the intensity vs energy transfer plot for the averaged ($\frac{1}{2}$,$\frac{1}{2}$,$\frac{1}{2}$) (on-peak) positions compared to other averaged (off-peak) positions at a temperature of 80 mK.}]{High-resolution neutron scattering data of Tb$_2$Ti$_2$O$_7$ in zero field. Panel (a) shows a plot of intensity vs energy transfer of the ($\frac{1}{2}$,$\frac{1}{2}$,$\frac{1}{2}$) peak intensity averaged over two peaks as a function of temperature. Note the logarithmic intensity scale. The ${\bf Q}$-integration range (H$=[\pm0.3,\pm0.7]$, L~$=[1.2,1.8]$ r.l.u.) is shown in the inset by the two black squares. Panel (b) shows the intensity vs energy transfer plot for the averaged ($\frac{1}{2}$,$\frac{1}{2}$,$\frac{1}{2}$) (on-peak) positions compared to other averaged (off-peak) positions at a temperature of 80 mK. The integration ranges for the averaging and extraction of the energy dependence is depicted in the inset by the black and magenta squares.}
\label{fig:figure5b6}
\end{figure}

Higher energy resolution neutron scattering measurements with $E_i=1.28$ meV were performed to investigate the temperature and ${\bf Q}$-dependence of the low-lying spin excitations in Tb$_2$Ti$_2$O$_7$ in zero magnetic field.  Figure \ref{fig:figure5b6}(a) shows a plot of intensity vs energy transfer obtained by integrating over the two ($\frac{1}{2}$,$\frac{1}{2}$,$\frac{1}{2}$) positions in reciprocal space shown as the black squares in the inset to Fig. \ref{fig:figure5b6}(a). Data is shown for three different temperatures, in the ground state ($T=80$ mK), near the spin glass transition ($T=275$ mK) and well above the transition ($T=600$ mK). As the temperature is lowered through the transition at 275 mK, spectral weight is transferred from the low-energy inelastic channel into the resolution-limited elastic channel (-0.02 meV $<E<$0.02 meV), opening up a gap of $\sim0.06-0.08$ meV. The low-lying spin excitation spectrum above the gap extends $\sim$0.2meV. This collapse of the magnetic spectral weight over much of the magnetic Brillouin zone into the elastic channel is a canonical signature of spin freezing behavior, as observed in spin glasses such as, for example, Y$_2$Mo$_2$O$_7$.\cite{JasonY2Mo2O7}

$\mu$SR measurements\cite{Gardner1999,Yaouanc2011} in zero field have consistently reported persistent dynamics in the spin lattice relaxation rate down to very low temperatures in Tb$_2$Ti$_2$O$_7$, which may suggest that either the elastic scattering itself is not truly elastic on the lower frequency scale probed by $\mu$SR, or that the depletion of spectral weight in the spin gap is not complete. 

Weak {\bf Q}-dependence is observed to the inelastic magnetic spectral weight at low temperatures, as shown in Fig. \ref{fig:figure5b6}(b). This plot compares intensity vs energy transfer over small regions of reciprocal space centered on {\bf Q}=$\left(\frac{1}{2},\frac{1}{2},\frac{1}{2}\right)$ wave vectors, and positions in reciprocal space far-removed from $\left(\frac{1}{2},\frac{1}{2},\frac{1}{2}\right)$ positions (referred to in Fig. \ref{fig:figure5b6}(b) as ``on-peak" and ``off-peak" positions, which are shown in the black and magenta boxes, respectively, in the inset to Fig. \ref{fig:figure5b6}(b)).  The low-energy inelastic scattering appears to be gapped over the entire Brillouin zone, the spectral weight extends $\sim0.03$ meV lower in energy at the {\bf Q}=$\left(\frac{1}{2},\frac{1}{2},\frac{1}{2}\right)$ frozen spin ice zone center as compared with {\bf Q}-positions in the middle of the zone. While Tb$_2$Ti$_2$O$_7$ in $H=0$ does not display conventional long-range order at any temperature and therefore does not exhibit well defined spin wave excitations, this observation is consistent with a dispersion of a sort in the small lifetime spin excitations, softening somewhat at the {\bf Q}=$\left(\frac{1}{2},\frac{1}{2},\frac{1}{2}\right)$ frozen spin ice zone center.

Taken together, our new neutron scattering data is fully consistent with a spin glass transition near $T_{\rm g}\sim 275$ mK in $H=0$. This (cluster) spin glass state builds upon short-range antiferromagnetic \textit{spin ice} correlations, as opposed to either primarily short-range antiferromagnetic (e.g. La$_{2-x}$Sr$_x$CuO$_4$\cite{Fujita2002}) or ferromagnetic correlations (e.g. Eu$_x$Sr$_{1-x}$S\cite{Maletta1983}), and has a low temperature isotropic correlation length of $\sim8$ \AA, as previously reported. The collapse of the low energy magnetic spectral weight across the magnetic zone into short-range ordered elastic scattering at {\bf Q}=$\left(\frac{1}{2},\frac{1}{2},\frac{1}{2}\right)$ wave vectors is a necessary component to this picture, informing on the magnetism in the frequency or time domain. It is also inconsistent with the singlet ground state scenario proposed theoretically for Tb$_2$Ti$_2$O$_7$ at low temperatures, as the inelastic magnetic intensity which forms the gap is transferred to elastic magnetic scattering, as opposed to being transferred above the gap leaving a nonmagnetic ground state and an absence of elastic magnetic scattering. The magnetic spin glass ground state based on antiferromagnetic spin ice is also consistent with ac-susceptibility measurements which also show canonical signatures for spin glass ground states in a variety of Tb$_2$Ti$_2$O$_7$ samples, both in polycrystalline and single crystal form. While there is some sample variation observed in the precise temperature at which a signature for a spin glass state is seen in Tb$_2$Ti$_2$O$_7$, all samples reported to date show such a signature in the approximate temperature range of 150-275 mK, consistent with the neutron scattering measurements we present in this work.

\section{Specific Heat}

Significant sample variability has, however, been reported in anomalies, and the lack thereof, in the low temperature heat capacity measured on different samples of Tb$_2$Ti$_2$O$_7$. We note that sample variability in the position and sharpness of heat capacity anomalies at low temperature in Yb$_2$Ti$_2$O$_7$ have also been reported,\cite{Ross2012stuffedYTO} and these have been ascribed to the relative role of weak ``stuffing" (excess Yb ions substituting on the Ti sublattice at the $\sim 2$\% level). In particular, Taniguchi and co-workers studied a family of polycrystalline samples of the form Tb$_{2+x}$Ti$_{2-x}$O$_7$ for very small non-stoichiomtery, $|x|<0.005$.\cite{Taniguchi2013} They report very significant variability in both the temperature at which the heat capacity anomaly is observed, as well as in the amplitude of the anomaly. These anomalies in $C_{\rm p}$ are typically observed in the 400 mK to 500 mK regime in zero field, and these results present two important questions for our study. First, does our neutron scattering single crystal sample of Tb$_2$Ti$_2$O$_7$ display such a $C_{\rm p}$ anomaly; if so how strong and at what temperature? Second, what is the relation between these sample dependent $C_{\rm p}$ anomalies in zero field in Tb$_2$Ti$_2$O$_7$ and the transition to the frozen spin ice state we observe in our neutron scattering sample at $T_{\rm g}\sim275$ mK?

\begin{figure}[h]
\includegraphics[width=8cm]{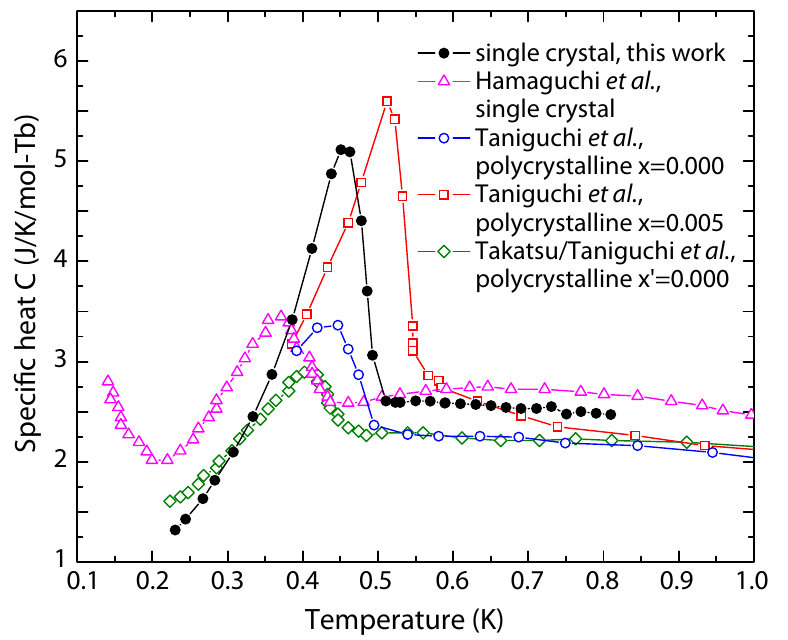}%
\caption{Specific heat versus temperature for different samples of Tb$_2$Ti$_2$O$_7$ in zero magnetic field. Measurements from a single crystal cut from the neutron scattering sample which we report on, is shown in black.  The feature at 450 mK does not coincide with the glass-like transition at 275 mK indicated by our neutron scattering measurements. For comparison, we also show $C_{\rm p}$ data taken from the literature.\cite{Hamaguchi2004,Taniguchi2013}}
\label{fig:figure5b7}%
\end{figure}

To address these questions we performed specific heat measurements down to a temperature of 220 mK on a 33.8 mg single crystal sample, cut from the neutron scattering sample which was the focus of this paper and, as already mentioned, that of a previous study.\cite{Rule2006} Figure \ref{fig:figure5b7} shows the measured specific heat of our single crystal sample of Tb$_2$Ti$_2$O$_7$, compared with several other samples, both single crystals and polycrystals, taken from the literature.\cite{Hamaguchi2004,Taniguchi2013} As can be seen from Fig. \ref{fig:figure5b7}, we observe a relatively large anomaly in $C_{\rm p}$ at $\sim450$ mK, almost twice the $T_{\rm g}$ associated with the frozen spin ice state characterized in this paper with neutron scattering. For reference, we have indicated the location of the $C_{\rm p}$ anomaly in zero field in Fig. \ref{fig:figure5b3}(c), which also shows the temperature dependence of the {\bf Q}$=\left(\frac{1}{2},\frac{1}{2},\frac{1}{2}\right)$ elastic scattering. Clearly, the temperature scales associated with the results of these measurements are distinct. As this $\sim450$ mK temperature scale is not present in either our neutron scattering results or ac-susceptibility results on Tb$_2$Ti$_2$O$_7$ taken from the literature, we conclude that while this heat capacity anomaly and its associated sample dependence are intriguing and potentially important phenomena, they do not seem to originate from magnetic dipole degrees of freedom in Tb$_2$Ti$_2$O$_7$, similarly to what has recently been suggested in Yb$_2$Ti$_2$O$_7$.\cite{DOrtenzio2013}

\section{Conclusion}

In conclusion, our neutron scattering data on Tb$_2$Ti$_2$O$_7$ in a magnetic field show that a small field of $\sim$ 0.075 T applied along [1-10] is sufficient to destroy the short-range antiferromagnetically ordered spin ice state characterized by diffuse elastic scattering at ($\frac{1}{2}$,$\frac{1}{2}$,$\frac{1}{2}$) positions. This field range is consistent with a $\mu_0H_{\rm C}\sim0.067$ T for fields along [111] reported by Yin \textit{et al.}, \cite{Yin2013} who also suggest that the zero-field ground state of Tb$_2$Ti$_2$O$_7$ is a quantum spin ice. The temperature dependence of the elastic diffuse scattering reveals a spin glass-like freezing at $T_{\rm g}\sim275$ mK. The details of the neutron signatures associated with the ($\frac{1}{2}$,$\frac{1}{2}$,$\frac{1}{2}$) elastic scattering display some history-dependence. The appearance of the elastic diffuse scattering below 275 mK correlates with the opening of a spin gap of 0.06-0.08 meV over most of reciprocal space and concomitant weakly dispersive spin excitations extending out to $\sim$0.2 meV. These excitations appear to go soft at the short-range ordering positions of ($\frac{1}{2}$,$\frac{1}{2}$,$\frac{1}{2}$). Specific heat measurements on a piece of the same single crystal show an anomaly at 450 mK, consistent with other reports by Hamaguchi \textit{et al.} and Taniguchi \textit{et al.}. We observe no signs for an anomaly near 275 mK, indicating that different degrees of freedom are involved in the ordering process responsible for this specific heat anomaly in Tb$_2$Ti$_2$O$_7$. The new neutron measurements we report on the temperature and magnetic field parametrics allow us to complete the $H$-$T$ magnetic phase diagram for Tb$_2$Ti$_2$O$_7$ in a [1-10] magnetic field.  We find the dynamic and disordered, cooperative paramagnetic phase, which generated much of the original interest in Tb$_2$Ti$_2$O$_7$, to be bracketed by a low-field frozen spin ice state, and a field-induced, long-range ordered antiferromagnetic state at high fields.

\section*{Acknowledgments}
The authors acknowledge useful contributions from M. J. P. Gingras. This work utilized facilities supported in part by the National Science Foundation under Agreement No. DMR-0944772, and was supported by NSERC of Canada. The DAVE software package\cite{DAVE} was used for data reduction and analysis of DCS data.

\end{document}